\documentclass{PoSa}

\usepackage{amsmath,microtype,multirow}
\newcommand{\EeV}{\mathrm{EeV}}

\title{Full-sky searches for anisotropies in UHECR arrival directions with the Pierre Auger Observatory and the Telescope Array}

\ShortTitle{Full-sky UHECR anisotropy searches with Auger and TA}

\author{
	\speaker{A. di Matteo}$^a$,
	T.~Bister, J.~Biteau, L.~Caccianiga, O.~Deligny, T.~Fujii, D.~Harari,
	D.~Ivanov, K.~Kawata, J.P.~Lundquist, R.~Menezes, D.~Mockler, T.~Nonaka,
	H.~Sagawa, P.~Tinyakov, I.~Tkachev, and S.~Troitsky,
	for the Pierre Auger Collaboration$^b$\footnote{for collaboration lists see \protect\pos{PoS(ICRC2019)1177}}~
	and the Telescope Array Collaboration$^c$\footnotemark[2]\\
	\llap{$^a$}Istituto Nazionale di Fisica Nucleare (INFN), Sezione di Torino, Turin, Italy\\
	\llap{$^b$}Observatorio Pierre Auger, Av.\ San Mart\'in Norte 304, 5613 Malarg\"ue, Argentina\\
	E-mail: \href{mailto:auger_spokespersons@fnal.gov}{\rm auger\_spokespersons@fnal.gov}\\
	Full author list: \href{http://www.auger.org/archive/authors_icrc_2019.html}{\rm http://www.auger.org/archive/authors\_icrc\_2019.html}\\
	\llap{$^c$}Telescope Array Project, 201 James Fletcher Bldg, 115 S.\ 1400 East, Salt Lake City, UT 84112-0830, USA\\
	E-mail: \href{mailto:ta-icrc@cosmic.utah.edu}{\rm ta-icrc@cosmic.utah.edu}\\
	Full author list: \href{http://www.telescopearray.org/research/collaborators}{\rm http://www.telescopearray.org/research/collaborators}
}

\abstract{The arrival directions of ultra-high-energy cosmic rays appear to be approximately isotropically distributed over the whole sky, but the last-generation UHECR detector arrays, the Pierre Auger Observatory (Auger) and the Telescope Array (TA), have detected thousands of events, allowing us to study small deviations from isotropy.  So far, Auger has detected a large-scale anisotropy consistent with a $\sim 6.5\%$ dipole moment in the distribution of cosmic rays with $E > 8$~EeV, and both collaborations have reported indications for smaller-scale anisotropies at higher energies.  On the other hand, neither array has full-sky coverage, the Auger field of view being limited to declinations $\delta < +45^\circ$ and the TA one to $\delta > -16^\circ$.  Searches for anisotropies with full-sky coverage thus require combining data from both arrays.  A working group with members from both collaborations has been established for this task.  Since even a minor systematic error in the energy determinations at either of the arrays could result in a sizeable spurious anisotropy along the north--south axis, we devised a method to cross-calibrate the energy scales of the two experiments with respect to each other by using events in a declination band within the intersection of their fields of view.  In this contribution, we report on both updates on blind searches for anisotropies and the first full-sky studies of flux models based on possible local extragalactic sources.}

\FullConference{36th International Cosmic Ray Conference --- ICRC2019\\
                24 July -- 1 August, 2019\\
                Madison, Wisconsin, USA}

\begin{document}
\section{Introduction}
\subsection{Motivation and recent results}
\begin{figure}[t]
	\centering
	\includegraphics{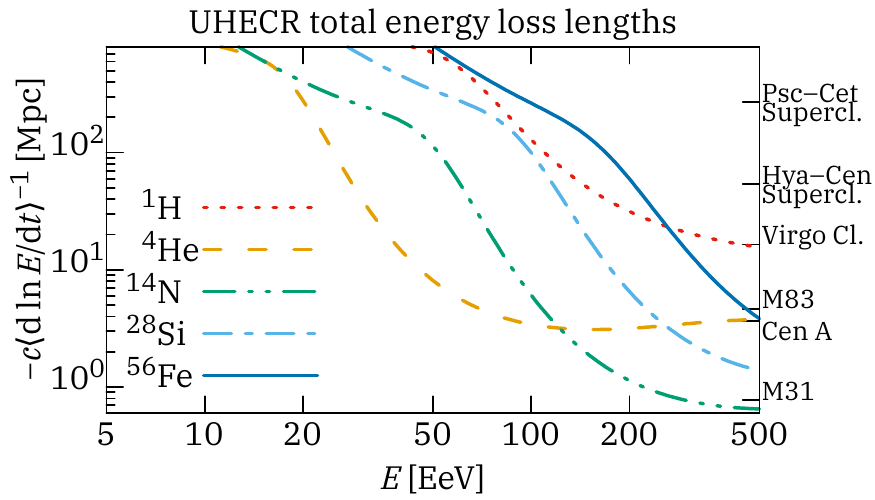}
	\caption{Energy loss lengths by UHECRs in the cosmic microwave background and extragalactic background light at $z=0$
		(computed by \textit{SimProp}~v2r4 \cite{SimProp} using the TALYS photodisintegration model \cite{TALYS} and the G12 EBL model \cite{Gilmore}),
		compared to the distance to a few selected nearby large-scale structures (Pisces--Cetus Supercluster, Hydra--Centaurus Supercluster, Virgo Cluster)
		and galaxies in the Local Sheet (M83, Cen~A, M31)}
	\label{fig:lambda}
\end{figure}

The origin of ultra-high-energy cosmic rays (UHECRs) is still unknown,
but at the highest energies their propagation distances are limited to a few tens of megaparsecs
by interactions with background photons (fig.~\ref{fig:lambda}),
and the Universe is not homogeneous at these scales.
Therefore, the distribution of their arrival directions can be expected to reflect the large-scale structure of the local Universe.
On the other hand, deflections by extragalactic and Galactic magnetic fields
(of order $30Z(E/10~\EeV)^{-1}$~degrees for nuclei with energy $E$ and atomic number $Z$)
can blur and distort the picture.
Anisotropies are expected to be stronger at higher energies, because both the propagation horizon and the magnetic deflections are reduced,
but so are the UHECR fluxes and consequently the statistical sensitivity of the available data.
Theoretical predictions of the level of anisotropy expected can be found e.g.\ in Ref.~\cite{predictions} and references therein.

The observed distribution of UHECR arrival directions appears to be relatively close to isotropic:
the only deviation observed with high statistical significance so far is a first harmonic in right ascension at energies above $8~\EeV$
that can be interpreted as a dipole moment of $6.5^{+1.3}_{-0.9}\%$ if higher-order moments are negligible,
which required over 30\,000 events to detect \cite{dipole}.  This dipole moment appears to grow proportionally to $E^{0.8\pm0.2}$,
in agreement with theoretical predictions assuming a mixed mass composition.
The lack of detectable anisotropies at energies from 1~to 3~EeV has been used to set stringent upper limits to the fraction of UHECRs that can be protons of Galactic origin \cite{Auger-limits,gal-limits}.
At higher energies ($E \ge 39~\EeV$),  there is an indication of a correlation with the positions of nearby starburst galaxies \cite{SBG-correl,TA-correl}.
At even higher energies, it has been reported that there is a region  of the sky in the northern hemisphere with a deficit of events up to $10^{19.75}$~eV and an excess of events with higher energies \cite{hotspot}.

\subsection{Datasets used in this work}
The largest cosmic ray detector arrays currently in operation are the Pierre Auger Observatory (Auger) \cite{Auger}
and the Telescope Array (TA) \cite{TA}.
Auger is located in the Mendoza Province, Agentina, at $35.2^\circ$~S, $69.2^\circ$~W, 1400~m above sea level.
Its main array consists of 1600 water Cherenkov detectors in a 1.5~km triangular grid
(covering about 3000~km$^2$ in total) and has been taking data since January~2004\@.
TA is located in Millard County, Utah, USA, at $39.3^\circ$~N, $112.9^\circ$~W, 1400~m above sea level.
Its main array consists of 507 plastic scintillator detectors in a 1.2~km square grid
(covering about 700~km$^2$ in total) and has been taking data since May~2008\@.
In this work, we use the same datasets as in Ref.~\cite{UHECR18}, where more details about the detectors, the selection criteria and the cross-calibration procedure can be found.

\begin{figure}[t]
	\centering
	\includegraphics[page=1]{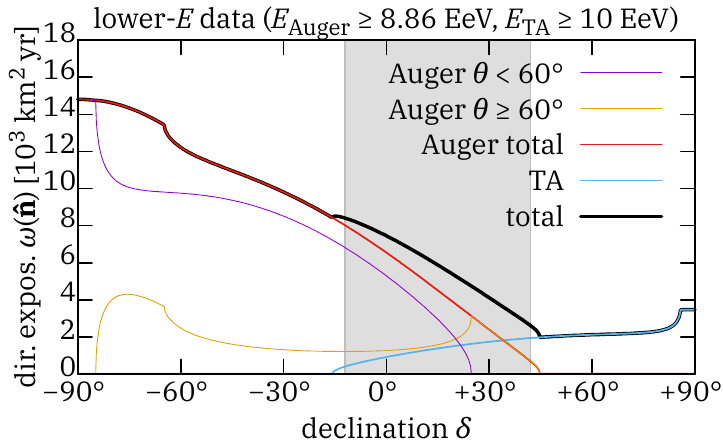}\hfil\includegraphics[page=2]{exposures.pdf}
	\caption{Directional exposures of the datasets used in this work.  The thresholds were chosen to match the integral fluxes in the shaded band (see Ref.~\cite{UHECR18} for details).}
	\label{fig:exposures}
\end{figure}

The lower-energy dataset includes events detected
by TA from 2008 May~11 to 2017 May~10 with reconstructed energies $E \ge 10~\EeV$ and zenith angles $\theta < 55^\circ$,
and events detected by Auger from 2004 Jan~01 to 2016 Aug~31 with $E \ge 8.86~\EeV$ and $\theta < 80^\circ$,
with different quality cuts and reconstruction techniques for $\theta \lessgtr 60^\circ$.
The higher-energy dataset includes TA events in the same time period with $E \ge 53.2~\EeV$
and Auger events until 2017 Apr~30 with $E \ge 40~\EeV$.
These energy thresholds were chosen in order to match the computed integral fluxes in the declination band $-12^\circ \le \delta \le +42^\circ$,
which is visible to both detectors, as described in Ref.~\cite{UHECR18}, because due to the steeply decreasing spectrum at these energies even a minor mismatch in the actual energy thresholds of the Auger and TA data would have resulted in a sizeable mismatch in the corresponding integral fluxes and hence a spurious reconstructed north--south anisotropy.
The mismatches in nominal energies are within the stated systematic uncertainties on the energy scales
($\pm 14\%$ for Auger, $\pm 21\%$ for TA) and in good agreement with those found by the Auger--TA energy spectrum working group \cite{spectrum} only using declinations $\delta \le +24.8^\circ$ and Auger events with $\theta < 60^\circ$.
The detectors are assumed to be fully efficient in these energy ranges,
i.e.\ their exposures are approximated as proportional to the geometrical exposures, as shown in fig.~\ref{fig:exposures}.

In total, we have about 31\,000 events in the lower-energy dataset and 969 events in the higher-energy dataset.

\section{Results}
\subsection{Search for full-sky large-scale anisotropies at relatively low energies}
\begin{figure}[t]
	\centering
	\includegraphics[width=2.7in]{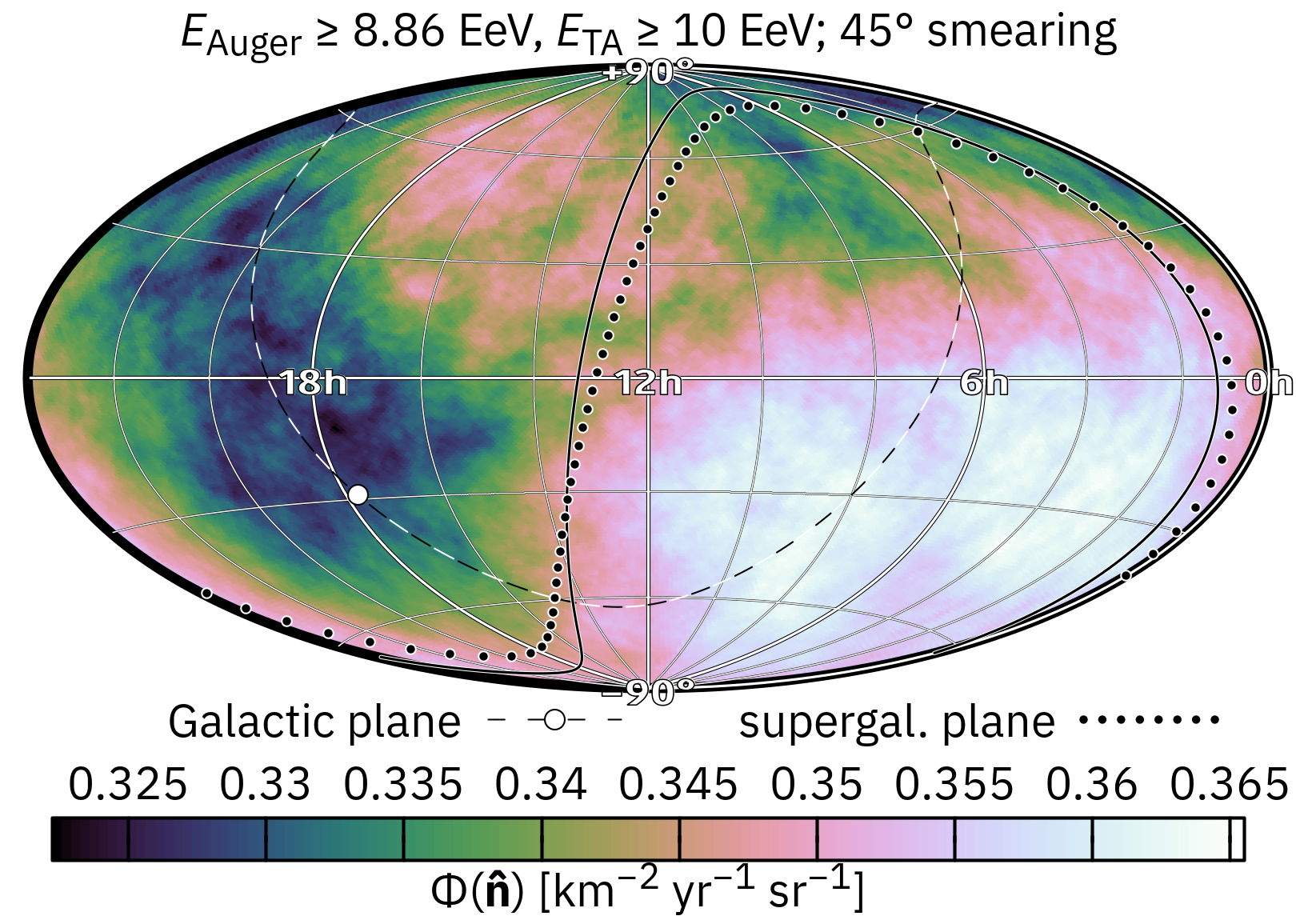}\hfil%
	\begin{tabular}[b]{cr@{ = }lc}
		\hline
		\multirow{3}{*}{This work}
		& $d_x$ & \multicolumn{2}{@{}l}{$(-0.7 \pm 1.1 \pm 0.01)\%$} \\
		& $d_y$ & \multicolumn{2}{@{}l}{$(+4.2 \pm 1.1 \pm 0.04)\%$} \\
		& $d_z$ & \multicolumn{2}{@{}l}{$(-2.6 \pm 1.3 \pm 1.4)\%$} \\
		\hline
		& $d_x$ & $(-1.0 \pm 1.0)\%$ & \multirow{3}{*}{$\ell_{\max} = 1$} \\
		& $d_y$ & $(+5.9 \pm 1.0)\%$ & \\
		Auger \cite{dipole} & $d_z$ & $(-2.6 \pm 1.5)\%$ & \\
		\cline{2-4}
		$\ge 8~\EeV$ & $d_x$ & $(-0.3 \pm 1.3)\%$ & \multirow{3}{*}{$\ell_{\max} = 2$} \\
		& $d_y$ & $(+5.0 \pm 1.3)\%$ & \\
		& $d_z$ & $(-2\phantom{.0} \pm 4)\%$ & \\
		\hline
	\end{tabular}
	\caption{Left:~Flux of cosmic rays from the lower-energy dataset smoothed in $45^\circ$-radius circular windows (Hammer projection, equatorial coordinates). Right:~Corresponding components of the dipolar moment in equatorial coordinates. The two uncertainties on results from this work are the statistical one and that resulting from the energy scale cross-calibration, respectively.  The two Auger-only reconstructions assumed the anisotropy to be purely dipolar ($\ell_{\max}=1$) or dipolar and quadrupolar ($\ell_{\max}=2$).}
	\label{fig:lowE}
\end{figure}

We used our lower-energy dataset to compute estimators of the dipole moment components which are unbiased regardless of any higher multipole moments, as explained in Ref.~\cite{UHECR18}.
This procedure would not have been possible using only one detector, because it requires non-vanishing directional exposure over the whole celestial sphere.
Our results are shown in fig.~\ref{fig:lowE}.
Previous results had been obtained using Auger data only \cite{dipole} with a slightly lower energy threshold, with which the current results are in good agreement, but the latter have substantially smaller uncertainties than the former when allowing for possible non-vanishing quadrupole moments, especially on the north--south component.
The uncertainty on the north--south dipole component in the current results is mostly due to that on the cross-calibration of energy scales.  The dipole moment we reconstructed corresponds to an angular power spectrum coefficient $C_1 = 3.5\times10^{-3}$ (normalized to $C_0 = 4\pi$), only exceeded in 1.3\% of isotropic simulations.\footnote{In each simulation, the exposure ratio between the experiments was allowed to fluctuate according to the uncertainties in the cross-calibration of energy scales.  When keeping the exposure ratio fixed, only 0.05\% of isotropic simulations had a higher $C_1$ than the data.}

The flux pattern visible in fig.~\ref{fig:lowE} does not look purely dipolar: a large brightest region is clearly visible in the south-east, but in the north-west rather than just a large darkest region there also seems to be a second relatively bright region.
This may interpreted as an indication for a possible quadrupole moment: the reconstructed angular power spectrum coefficient $C_2 = 1.3\times 10^{-3}$ was only exceeded in 5.5\% of isotropic simulations.\footnote{4.2\% when keeping the exposure ratio fixed (see previous footnote).}  In the future we will try to confirm or refute the presence of a quadrupolar pattern using more data.

\subsection{Blind search for medium-scale anisotropies at the highest energies}
We computed the unbiased estimator of the integral flux (see Ref.~\cite{UHECR18} for details) above the higher energy thresholds
in circular windows of $5^\circ$, $10^\circ$, \ldots, and $35^\circ$ radius, and the corresponding local Li--Ma significances
against the isotropic null hypothesis.
\begin{figure}[t]
	\centering
	\includegraphics[width=2.7in]{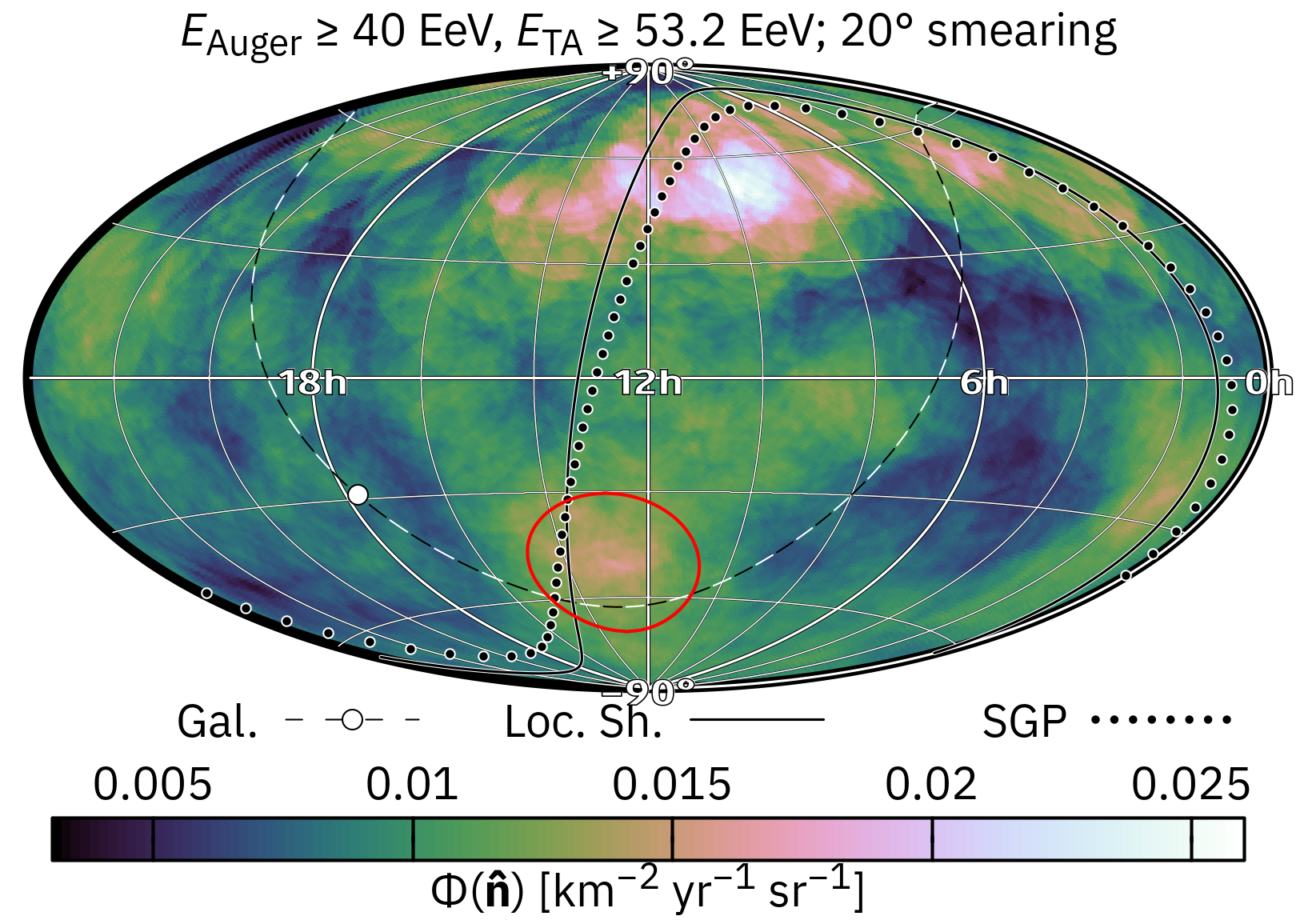}\hfil\includegraphics[width=2.7in]{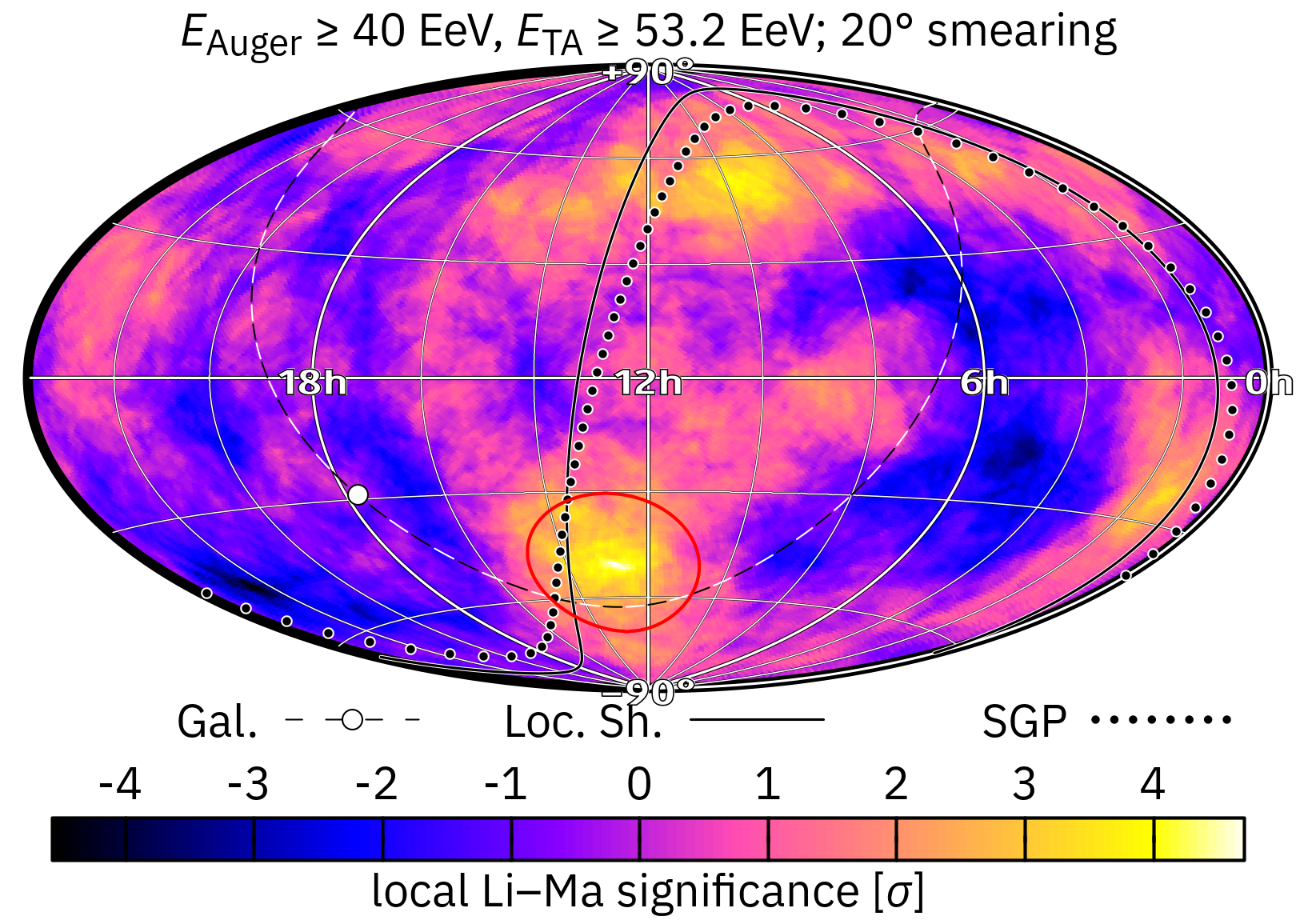}
	\includegraphics[width=2.7in]{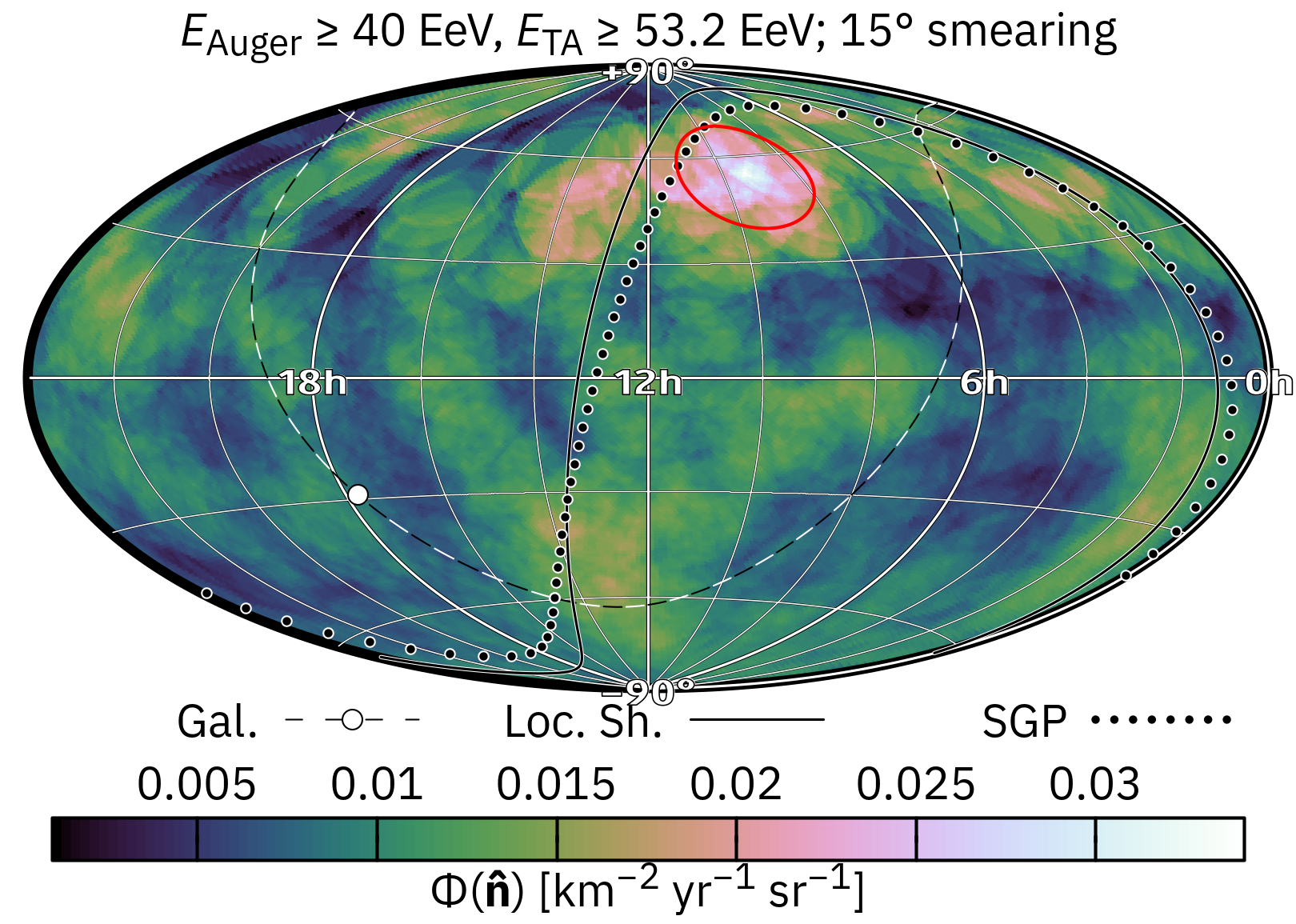}\hfil\includegraphics[width=2.7in]{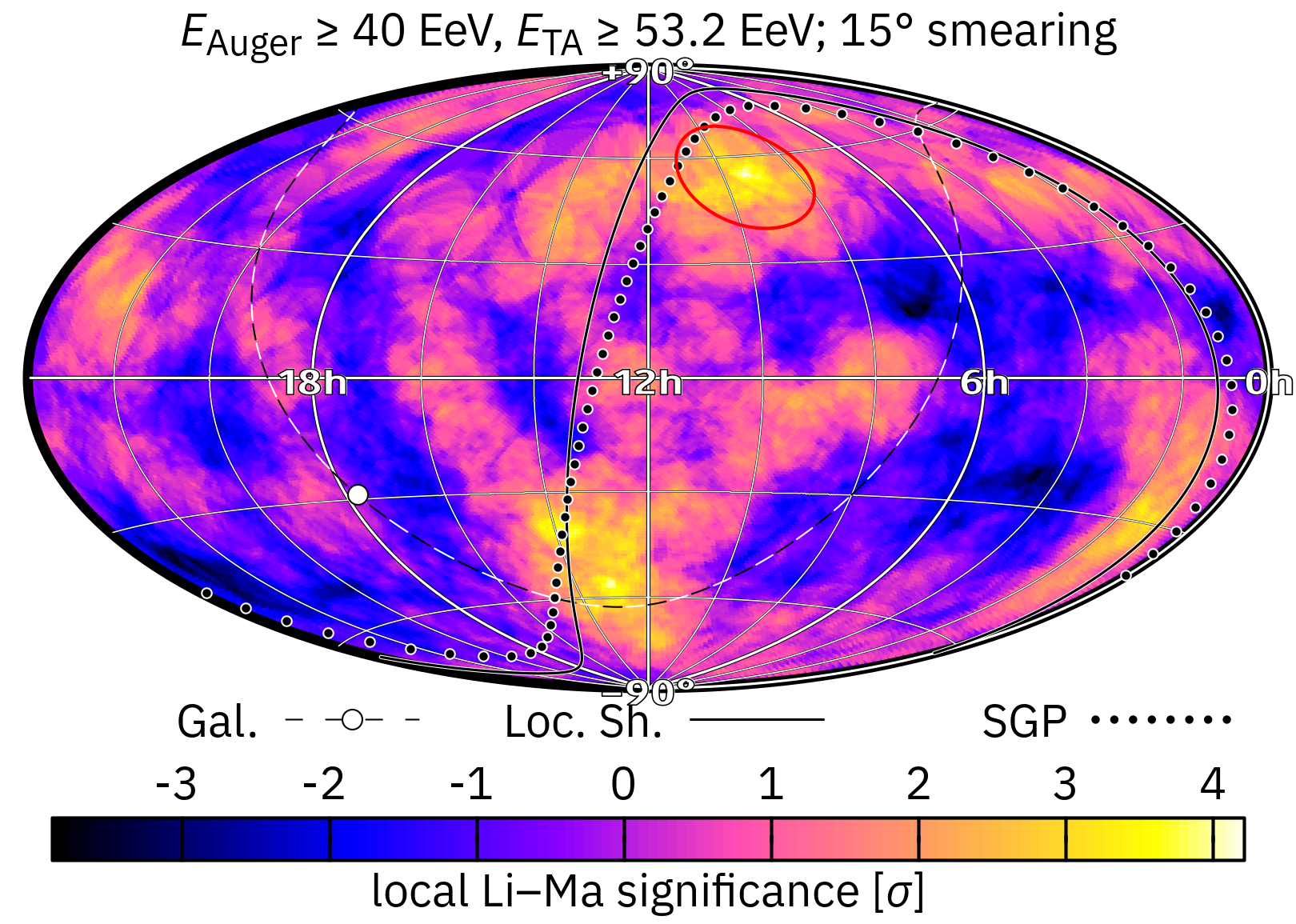}
	\caption{Left:~Flux of cosmic rays from the higher-energy dataset smoothed in  $20^\circ$-radius (top) and $15^\circ$-radius (bottom) circular windows (Hammer projection, equatorial coordinates),
	with the most significant excess highlighted (red circle).
		Right:~Corresponding local statistical significances against the isotropic null hypothesis.}
	\label{fig:highE}
\end{figure}
The two most significant excesses are found
in a $20^\circ$-radius window around $(\alpha=12^\mathrm{h~}50^\mathrm{m}, \delta=-50^\circ)$ with $4.7\sigma$ local significance, and
in a $15^\circ$-radius window around $(9^\mathrm{h~}30^\mathrm{m}, +54^\circ)$ with $4.2\sigma$ local significance,\footnote{
	This significance is not directly comparable to that in the TA-only study \cite{old-hotspot}, because in this work we are using a lower energy threshold (53.2 vs 57~EeV on the TA energy scale),
	and the deficit of events below $10^{19.75}$~eV \cite{hotspot} partially cancels out the excess at higher energies,
	resulting in a weaker overall anisotropy.
} shown in fig.~\ref{fig:highE}.
Taking into account the scan over window sizes and positions, these correspond to a $2.2\sigma$ and a $1.5\sigma$ post-trial significance respectively.

\subsection{Search for excesses along the supergalactic plane or the Local Sheet}
It has been pointed out that all the three most visible excesses in our higher-energy data (fig.~\ref{fig:highE}) are relatively close to the supergalactic plane,
a well-known structure along which galaxy clusters within several tens of Mpc of us are preferentially located, recognized by G.\ de~Vaucouleurs in 1953.
They are also close to certain galaxies in the Local Sheet \cite{Local-Sheet}, a planar structure about 0.5~Mpc thick and 10~Mpc across tilted by only $8^\circ$ with respect to the supergalactic plane
and including nearly all galaxies within about 6~Mpc of us:
namely, the excess region shown in fig.~\ref{fig:highE}~(top) includes NGC~4945, Cen~A and Circinus; M83 is within $3^\circ$ of its edge; and M81 and M82 are within~$1^\circ$ of the edge of the one in Fig.~\ref{fig:highE}~(bottom).
The Local Sheet consists of the Local Group (the Milky Way and M31, plus their satellites) near the center
and the Council of Giants (twelve more large galaxies and their satellites) in a ring surrounding it.
It includes several starburst galaxies and an AGN, so that most of the anisotropic UHECR flux in the best-fit SBG- and AGN-based models of Ref.~\cite{SBG-correl} originates from within the Local Sheet.

\begin{figure}[t]
	\centering
	\includegraphics[page=1]{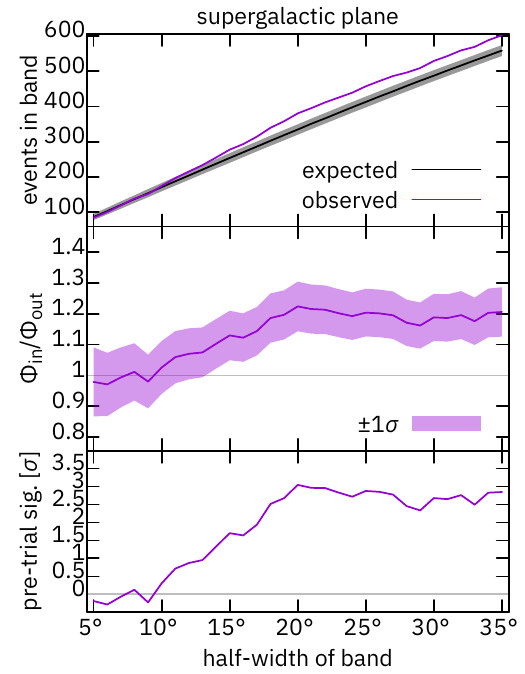}\hfil\includegraphics[page=2]{BandTest.pdf}
	\caption{
		Top: Number of events in our higher-energy dataset within $5^\circ, 6^\circ, \ldots, 35^\circ$,
		of the supergalactic plane (left) or Local Sheet (right), compared to isotropic expectations\@.
		Middle: Ratio between the reconstructed fluxes inside and outside each band\@.
		Bottom: Local Li--Ma significance of excesses in each band.}
	\label{fig:supergal}
\end{figure}

To assess the statistical significance of the alignment of excesses along the supergalactic plane or the Local Sheet,
we counted events in bands centered on the supergalactic plane or the Local Sheet of $\pm 5^\circ$, $\pm 6^\circ$, \ldots and $\pm 35^\circ$
angular half-widths, and compared the results to the expectations assuming an isotropic flux given the integrated directional exposure of the detectors in each band and a fixed total number of events in the full sky,
computing local Li--Ma significances.  The results are shown in fig.~\ref{fig:supergal}.
The highest local significance we find is $+3.6\sigma$, in a band of $\pm 24^\circ$ around the Local Sheet, in which $455$ events are observed while $400\pm15$ are expected,
whereas the highest significance in a band centered on the supergalactic plane is $+3.0\sigma$, with $20^\circ$ half-width, in which $380$ events are observed while $335\pm15$ are expected.
We found the uncertainty in the energy cross-calibration to have a negligible impact on this analysis, its contribution to the uncertainties in $\Phi_\text{in}/\Phi_\text{out}$ being around two orders of magnitude smaller than the statistical uncertainties.

\begin{figure}[t]
	\centering
	\includegraphics{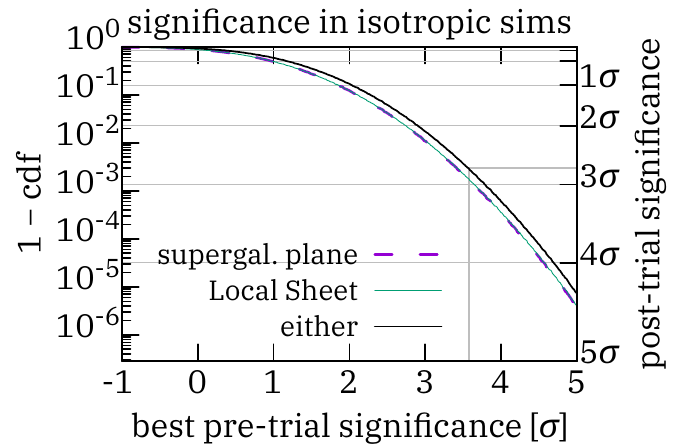}
	\caption{Cumulative distribution function (cdf) of best pre-trial significances in isotropic simulations and corresponding post-trial significances. The statistical penalty factor for using both the supergalactic plane and the Local Sheet over only using one is considerably less than 2,
because the two planes are only tilted by $8^\circ$ with respect to each other so the bands largely overlap.}
	\label{fig:cumulative}
\end{figure}

We performed the same scan over 31 band widths and two planes on over $2\times 10^8$ Monte Carlo datasets generated assuming an isotropic flux.  As shown in fig.~\ref{fig:cumulative}, we found a local significance greater than $3.6\sigma$ in $p=0.3\%$ of the simulations, corresponding to a $2.8\sigma$ global significance.

The events detected by Auger until 2014 March~31 ($62.1\%$ of the current dataset) were used in Ref.~\cite{Auger-scan}
for a similar study, in which the energy threshold was also allowed to vary between 40~EeV and 80~EeV in 1~EeV steps.
If the same energy scan were performed on the current data without finding any higher local significance,
the global $p$-value would increase to approximately $2\%$\@.  Other previous searches for supergalactic excesses in UHECR data include
Refs.~\cite{Stanev1,Stanev2,TA-supergal}.

\section{Possible future analyses}
In the present analysis, no attempt was made to explicitly use the information we have about positions of individual galaxies,
energy losses in UHECR propagation and the associated production of secondary particles,
or deflections in extragalactic and Galactic magnetic fields.
Doing so would allow us to define more physically motivated models of intermediate-scale anisotropies in the UHECR flux to be tested with our data.
As for large-scale anisotropies, in Ref.~\cite{predictions} it is predicted that our worst-case sensitivity to them would be maximized
by using an energy threshold intermediate between the two used in the present work.

In addition to this, upgrades of both the Pierre Auger Observatory and the Telescope Array are currently being deployed.
The former \cite{AugerPrime} will provide us with mass-sensitive observables on an event-by-event basis, allowing us to collect
event samples enriched in protons (which are guaranteed to have undergone relatively small magnetic deflections),
or in medium-light nuclei (which are guaranteed to have originated within relatively short distances).
The latter \cite{TAx4} will provide us with data with which to study anisotropies in the northern hemisphere at the highest energies
with greatly enlarged statistics.

\newcommand{\pub}[7]{#1, \href{https://doi.org/#6}{\textit{#2} \textbf{#3} (#4) #5} [\href{https://arxiv.org/abs/#7}{#7}]}
\newcommand{\pubnoarXiv}[6]{#1, \href{https://doi.org/#6}{\textit{#2} \textbf{#3} (#4) #5}}
\newcommand{\Auger}{A.\ Aab et al.\ [Pierre Auger Collaboration]}
\newcommand{\Augerold}{P.\ Abreu et al.\ [Pierre Auger Collaboration]}
\newcommand{\TA}{R.U.\ Abbasi et al.\ [Telescope Array Collaboration]}
\newcommand{\TAold}{T.\ Abu-Zayyad et al.\ [Telescope Array Collaboration]}
\newcommand{\arXiv}[1]{\href{https://arxiv.org/abs/#1}{arXiv:#1}}

\end{document}